\documentclass[aps,pra,preprint,superscriptaddress,showpacs,preprintnumbers,amsmath,amssymb]{revtex4-1}
\usepackage{graphicx}

\begin{document}

\preprint{}

\title{The Langevin-equation description of optomechanics with the dispersive and   dissipative optomechanical coupling}

\author{Alexander K. Tagantsev}
\email{alexander.tagantsev@epfl.ch}
\affiliation{Swiss Federal Institute of Technology (EPFL), School of Engineering, Institute of Materials Science, CH-1015 Lausanne, Switzerland}
\begin{abstract}
The description of the optomechanical systems is commonly based on the quantum Langevin equation formalism. This framework is introduced phenomenologically or based on a model Hamiltonian. 
However, once dealing with the optomechanical  Fabry-Perot cavity or the modified Michelson-Sagnac interferometer with a semitransparent mechanically active membrane inside, a model-free consideration is also possible by using an alternative approach. 
Such an approach, which is based on the classical wave equations, is popular in the gravitational-wave community where it is termed as  input-output relations approach. 
In this work, using the aforementioned approach, we derived the equations for the ladder operator of the intracavity field, stochastic back-action force, and the relation between the fields at the input mirror. 
Then we simplified the obtained results down to the range of applicability of the Langevin equation formalism and compared these with the corresponding predictions of the latter formalism. 
This enabled us to critically assess the validity of the Langevin equation formalism and rectify its range of applicability to find that the latter is more stringent than that intuitively expected.  
In the case where the dissipative optomechanical coupling is involved we identified appreciable problems with this formalism. It was found that, disregarding the fact that decay rate of the optomechanical Fabry-Perot cavity depends on its length, no dissipative optomechanical coupling is generated. 
This is in contrast with the prediction of the standard Langevin-equation based treatment.
It was found that the Langevin equation formalism misses a phase factor at the input field; this factor turns out to be important for the situation involving the dissipative optomechanical coupling.

\end{abstract}
\date{\today}
\maketitle
\section{Introduction}
Nowadays the cavity quantum optomechanics is considered to be an important branch  of quantum optics~\cite{Aspelmeyer2014}.
The principle workhorse of cavity optomechanics is the so-called \emph{dispersive coupling}, appearing due to a dependence of the cavity resonance frequency on the position of a mechanical oscillator.
In addition, quantum optomechanics deals with the so-called \emph{dissipative coupling}~\cite{Elste2009} and \emph{coherent coupling}~\cite{li2019a}.
The dissipative coupling  is due to the dependence of the cavity decay rate on the position of the mechanical oscillator while, in the case of the coherent coupling, the mechanical displacement brings about a coupling between two optical modes.
Once identified~\cite{Elste2009}, the dissipative coupling has been attracting considerable attention of theorists~\cite{huang2017,Weiss2013,Weiss2013a,Kilda2016,
vyatchanin2016,nazmiev2019,Vostrosablin2014,Tarabrin2013,Xuereb2011,Tagantsev2018,tagantsev2019,mehmood2019,Khalili2016,huang2018,huang2019gen,mehmood2018,dumont2019,tagantsev2020a,
tagantsev2020b,tagantsev2021,baraillon2020} and experimentalists~\cite{Li2009,Sawadsky2015,tsvirkun2015,Wu2014,meyer2016,zhang_2014}.
Importance of taking into account the dissipative coupling when describing the modified Michelson-Sagnac interferometer (MMSI) was demonstrated in Ref.\cite{Xuereb2011}. 
For the optical resonator with the membrane inside when it is close to the "symmetry-braking" point, it was shown that the constant of the dissipative coupling may exceed the typical values of the dispersive coupling constant \cite{tagantsev2020a}. 
Similarly, a strong dissipative coupling was predicted for an optical cavity with the membrane outside the cavity \cite{tagantsev2021}.   

The theoretical description of the dissipative optomechanical coupling commonly uses the  quantum Langevin equation formalism \cite{Elste2009}.
Hereafter, we will use L-formalism as a shorthand for the  quantum Langevin equation formalism.
This framework is based on a model Hamiltonian.
However, once dealing with relatively simple optomechanical systems, like the optomechanical Fabry-Perot cavity or MMSI~\cite{Xuereb2011,Sawadsky2015,Khalili2016} with a semitransparent mechanically active membrane inside, a model-free consideration is also possible by using an alternative approach.
Such an approach, which is based on the classical wave equations in the systems, is popular in the gravitational-wave community where it is termed as "input-output relations approach"~\cite{Buonanno2003,Danilishin2012,Khalili2016}.
Here a terminological remark is needed.
What we call L-formalism is also called "Input-output formulation"~\cite{Walls2008}.
We hope that our  use of the word combination "input-output relations approach" does not lead to any confusion.
Hereafter, we will use IO-formalism  as a shorthand for the input-output relations approach.

In the present paper, we give a comprehensive treatment of the above two systems using IO-formalism.
We simplified the obtained results down to the range of applicability of L-formalism and compared these with the results obtained with the latter formalism.
Assuming that the results of IO-formalism are fully trustable,
we critically assess the validity of L-formalism and rectify its range of applicability to find that the latter is more stringent than that intuitively expected. 
Based on obtained results we identify some inconsistencies of L-formalism,  mostly when it is dealing with the dissipative optomechanical coupling.
An important conclusion of the paper is that the dependence of the cavity decay rate on the coordinate of the mechanical oscillator does not necessarily imply the generation of the dissipative optomechanical coupling, which is conflict with  the result of the standard  L-formalism.  

In Sect.\ref{L} we summarize L-formalism in application to a one-mode optomechanical system. In Sect.\ref{INOUT}, the purely optical Fabry-Perot cavity is addressed using IO-formalism.
In Sect.\ref{OM} and \ref{Mi}, IO-formalism is applied to the optomechanical Fabry-Perot cavity and MMSI.
Section \ref{Di} is devoted to a discussion of the results obtained with the two methods.
Section \ref{Co} summarizes the paper.
\section{Langevin equation framework}
\label{L}
We consider an optical cavity excited with a strong monochromatic light of frequency  $\omega_\textrm{L}$, which is slightly detuned from one of the cavity resonance frequencies $\omega_c$.
We mean $|\Delta|\ll \omega_\textrm{FSR} $,   $\Delta =\omega_\textrm{L}-\omega_c$, $\omega_\textrm{FSR}= \pi c/L$, and $\omega_\textrm{FSR}$ where $\omega_\textrm{FSR}$, $L$, and $c$  are the free spectral range of the cavity, the cavity length, and the speed of light, respectively.

It is a high-fineness cavity, i.e. its linewidth $\gamma$ is small compared to $\omega_\textrm{FSR}$:
\begin{equation}
\label{Cond3}
\gamma \ll \omega_{\textrm{FSR}}.
\end{equation}
The cavity is also driven by delta-correlated vacuum noise.
The intra-cavity field is described by the Bose ladder operator $\textbf{a}$.
The light in the cavity is coupled to a mechanical oscillator of mass $m$, having the resonance frequency  $\omega_m$.
The position of the oscillator $x$ corresponds to an operator $\textbf{x} = x_{\textrm{zpf}}(\textbf{b}^\dag+\textbf{b})$, where $x_{\textrm{zpf}}=\sqrt{\frac{\hbar}{2m\omega_{\textrm{m}}}}$, $\textbf{b}$ being the Bose ladder operator.

In the Langevin equation framework, the description of such a system starts from the Hamiltonian~\cite{Elste2009}
\begin{equation}
\label{Ham}
\textbf{H}=\hbar\omega_{\textrm{c}}\textbf{a}^\dag\textbf{a}+
\hbar\omega_{\textrm{m}}\textbf{b}^\dag\textbf{b}-\hbar g_\omega\textbf{a}^\dag\textbf{a}(\textbf{b}^\dag+\textbf{b})-
i\hbar\sqrt{\frac{\gamma}{2\pi\rho}} \sum\nolimits_{q}(\textbf{a}^\dag\textbf{c}_q-\textbf{c}_q^\dag\textbf{a})\left[1-g_\gamma(\textbf{b}^\dag+\textbf{b})/\gamma\right]
\end{equation}
where, in addition to the common \emph{dispersive} optomechanical interaction controlled by the coupling constant
\begin{equation}
\label{go}
g_\omega=-(d\omega_c/dx)x_{\textrm{zpf}},
\end{equation}
the \emph{dissipative} optomechanical interaction controlled by the coupling constant
\begin{equation}
\label{gg}
g_\gamma=-(d\gamma/dx)x_{\textrm{zpf}}/2
\end{equation}
is introduced.
Here  $\textbf{c}_q$ are ladder operators for the electromagnetic bath (the bath Hamiltonian is omitted). $\rho$ is the density of states [per energy] of the bath.

In the presence of the strong driving field, in the reference frame rotating with frequency $\omega_L$,  starting from (\ref{Ham}), Elste et al~\cite{Elste2009} derived the following equations for the intra-cavity field
\begin{equation}
\label{aM}
\left[\frac{\gamma}{2}-i(\Omega+\Delta)\right] \textbf{a}(\Omega)
 =\sqrt{\gamma}\textbf{A}_{\textrm{in}}(\Omega)+
\left[ ig_\omega a_0 + g_\gamma(a_0- A_0/\sqrt{\gamma}) \right ]\frac{\textbf{x}(\Omega)}{x_{\textrm{zpf}}}
\end{equation}
where $\mathbf{a}(\Omega)$, $\textbf{x}(\Omega)$, and $\textbf{A}_{\textrm{in}}(\Omega)$ are the Fourier transforms of $\mathbf{a}$, $\textbf{x}$, and the operator of a flux-normalized input noise field, respectively. $A_0$ is the flux-normalized complex amplitude of the driving field and  $a_0$ is  a number-of-quantum-normalized complex amplitude of driving field inside the cavity.
Here, like in Ref.~\cite{Elste2009}, for simplicity we neglect the Lamb shift of the optical frequency.

The Fourier transform of operator of the backaction force acting on the mechanical oscillator was found in the form ~\cite{Elste2009} 
\begin{equation}
\label{F}
\textbf{F}(\Omega)=\frac{\hbar g_\omega}{x_{\textrm{zpf}}} a_0^*\textbf{a}(\Omega)
+i\frac{\hbar g_\gamma}{x_{\textrm{zpf}}}\frac{a_0^*\textbf{A}_{\textrm{in}}(\Omega)-A_0^*\textbf{a}(\Omega)}{\sqrt{\gamma}} + \tilde{\textrm{h}}.\textrm{c}.
\end{equation}
Hereafter, we use the symbol $\tilde{\textrm{h}}.\textrm{c}$ to denote the Hermitian conjugate taken at $-\Omega$.

With  $\textbf{A}_{\textrm{in}}(\Omega)$  meeting the standard relations \cite{Walls2008,Tagantsev2018}
\begin{equation}
\label{comA}
[\textbf{A}_{\textrm{in}}(\Omega),\textbf{A}^\dag_{\textrm{in}}(\Omega')]=\delta(\Omega-\Omega'),
\end{equation}
\begin{equation}
\label{avA}
<\textbf{A}_{\textrm{in}}^\dag(\Omega)\textbf{A}_{\textrm{in}}(\Omega')>=0, \qquad <\textbf{A}_{\textrm{in}}(\Omega)\textbf{A}_{\textrm{in}}(\Omega')>=0
\end{equation}
where $<...>$ and $[...,...]$ denote the ensemble averaging and the commutator, respectively, Eqs. (\ref{aM}) and (\ref{F}) imply the following expression for the stochastic backaction force~\cite{Elste2009,footnote13}
\begin{equation}
\label{S}
S(\Omega)=\gamma\left(\frac{\hbar | a_0|}{x_{\textrm{zpf}}}\right)^2
\frac{g_\omega^2}
{(\Delta+\Omega)^2+\gamma^2/4 }\left(1 + \frac{g_\gamma}{g_\omega}\frac{2\Delta+\Omega}{\gamma}\right) ^2.
\end{equation}
Here we define $<\textbf{F}^\dag(\Omega)\textbf{F}(\Omega')> = S(\Omega)\delta(\Omega-\Omega')$.

The Langevin equation formalism also yields an equation linking the operators for the input and cavity fields with the operators of the output field $\textbf{A}_{\textrm{out}}(\Omega)$ \cite{Xuereb2011}:

\begin{equation}
\label{INOUT}
\textbf{A}_{\textrm{in}}(\Omega)+\textbf{A}_{\textrm{out}}(\Omega)=\sqrt{\gamma}\mathbf{a}(\Omega)-\frac{g_\gamma a_0}{\sqrt{\gamma}}\frac{\textbf{x}(\Omega)}{x_{\textrm{zpf}}}.
\end{equation}
\section{Optical Fabry-Perot cavity: input-output relations approach}
\label{O}
\begin{figure}
\includegraphics [width=0.7\columnwidth,clip=true, trim=0mm 0mm 0mm 0mm] {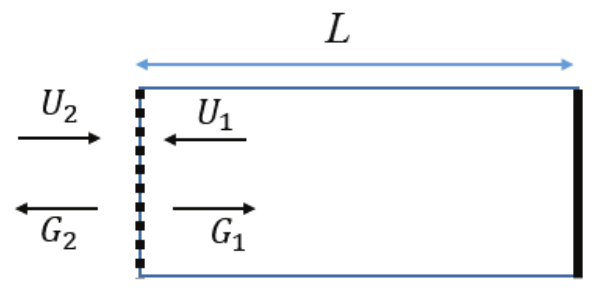}
\caption{Schematics of the one-sided Fabry-Perot cavity. The mirror shown with the dashed line is semitransparent, that shown with the solid line is perfectly reflecting. The running electromagnetic waves are schematically shown with the arrows and labeled  with their complex amplitudes.
\label{FPfig}}
\end{figure}

Let us consider the one-sided Fabry-Perot cavity, Fig.~\ref{FPfig}, as a practical realization of the system addressed in the previous section.
Now we use IO-formalism \cite{Buonanno2003,Danilishin2012,Khalili2016} for its description.
In this section, we consider the purely optical situation, where the mirrors of the cavity are immobile.

We characterize the input mirror with the following scattering matrix
\begin{equation}
\label{FPMatrix}
\mathbb{M}_{\textrm{FP}}= \left(
  \begin{array}{cc}
\rho &  \tau \\
\tau^* &-\rho^* \\
  \end{array}
\right) \qquad \rho=|\rho| e^{i\mu} \qquad \tau=|\tau| e^{i\nu},
\end{equation}
which is defined to relate the complex field amplitudes of the running waves (see Fig. \ref{FPfig}) as follows
\begin{equation}
\label{link}
\left(
  \begin{array}{cc}
G_1 \\
G_2  \\
  \end{array}
\right)=\mathbb{M}_{\textrm{FP}}
 \left(
  \begin{array}{cc}
U_1 \\
U_2  \\
  \end{array}
\right).
\end{equation}
We assume that the reflection against the backstop mirror occurs with the $\pi$ phase change of the light imposing an additional relation between the field amplitudes:
\begin{equation}
\label{ReflCond}
G_{1} =   - U_{1} e^{-2ikL}
\end{equation}
where $k$ is the wave vector of the light.

Following this approach, in the reference frame rotating with frequency $\omega_L$, we present the amplitudes of all the fields as the sum of a large constant amplitude of the light at the driving frequency and a small fluctuating part, i.e.
\begin{equation}
\label{twoparts}
  \begin{array}{cc}
 &  G_{1}(t)=e^{-i\omega_L t}[G_{10}+g_1(t)] \qquad G_{2}(t)=e^{-i\omega_L t}[G_{20}+g_2(t)] \\
 &  U_{1}(t)=e^{-i\omega_L t}[U_{10}+u_1(t)] \qquad U_{2}(t)=e^{-i\omega_L t}[U_{20}+u_2(t)]. \\
  \end{array}
\end{equation}
Next, Eqs.(\ref{FPMatrix}), (\ref{link}), and (\ref{ReflCond}) imply the following relationships for the Fourier transforms of fluctuating parts of the amplitudes
\begin{equation}
\label{Field1}
u_{1}(\Omega)(e^{-2iL(\Omega+\Delta)/c}-|\rho|)= |\tau| e^{i(\nu-\mu)} u_2(\Omega)
\end{equation}
\begin{equation}
\label{INOUT1}
e^{i\nu}g_{2}(\Omega)=|\tau| u_1(\Omega)-|\rho| e^{i(\nu-\mu)} u_2(\Omega)
\end{equation}
where the resonance condition $e^{-2i\omega_cL/c-i\mu}=-1$ was taken into account.

Now, to successfully pass from the classical equations (\ref{Field1}) and (\ref{INOUT1})  to the quantum equations, which are the most close to those of the Langevin framework, we apply the following correspondence rules
\begin{equation}
\label{rule1}
u_{1}(\Omega)\Rightarrow\sqrt{\frac{c}{2L}}\textbf{a}(\Omega)
\end{equation}
\begin{equation}
\label{rule2}
u_{2}(\Omega)\Rightarrow  e^{i(\mu-\nu)}\textbf{A}_{\textrm{in}}(\Omega).
\end{equation}
\begin{equation}
\label{rule11}
g_{2}(\Omega)\Rightarrow e^{-i\nu}\textbf{A}_{\textrm{out}}(\Omega),
\end{equation}
which provide us with the following quantum versions of Eqs. (\ref{Field1}) and (\ref{INOUT1}):
\begin{equation}
\label{a1}
\mathfrak{B}(\Omega) \textbf{a}(\Omega) =\sqrt{\gamma}\textbf{A}_{\textrm{in}}(\Omega)
\end{equation}
\begin{equation}
\label{INOUT2}
|\rho|\textbf{A}_{\textrm{in}}(\Omega)+\textbf{A}_{\textrm{out}}(\Omega)=\sqrt{\gamma}\mathbf{a}(\Omega).
\end{equation}
where
\begin{equation}
\label{B}
\mathfrak{B}(\Omega)= \frac{e^{-2iL(\Omega+\Delta)/c}-1}{2L/c}+\frac{\gamma}{2}\frac{2}{1+\sqrt{1-\tau^2}}
\end{equation}
and
\begin{equation}
\label{gamma}
\gamma=\frac{c|\tau|^2}{2L} .
\end{equation}

For sufficiently small $\gamma $, $\Omega$, and $\Delta$, one can replace $\mathfrak{B}(\Omega)$  with $-i(\Omega+\Delta)+\frac{\gamma}{2}$. 
One readily checks that, first, for  such a replacement the fulfillment of the following condition  
\begin{equation}
\label{Cond1}
\tau^2\ll 1
\end{equation}
 is necessary.  
This condition is consistent with the conclusion by Kampen~\cite{Kampen1997} that the Langevin equation is correct to within the lowest order in the coupling constant squared.

In addition, one needs that
\begin{equation}
\label{Cond2}
\frac{2L}{c}(\Omega+\Delta)^2\ll \gamma
\end{equation}
 and
\begin{equation}
\label{Cond22}
\frac{[2L(\Omega+\Delta)/c]^2}{3}\ll 1.
\end{equation}
The above three conditions can be summarized in the following two inequalities  
\begin{equation}
\label{CondF}
\gamma\ll \frac{\omega_{\textrm{FSR}}}{2\pi}
\end{equation}
and
\begin{equation}
\label{CondF1}
\gamma \gg \frac{2\pi(\Delta+\Omega)^2}{\omega_{\textrm{FSR}}}.
\end{equation}
One readily checks that, under conditions (\ref{CondF}) 
\begin{equation}
\label{Baprox}
\mathfrak{B}(\Omega)=\frac{\gamma}{2}-i(\Omega+\Delta)
\end{equation}
If inequalities (\ref{CondF}) and (\ref{CondF1}) are met,  $\mathfrak{B}(\Omega)$ can be approximated  with $\frac{\gamma}{2}-i(\Omega+\Delta)$ such that Eq.(\ref{a1}) leads to the standard Langevin equation for the purely optical system, c.f. Eq.(\ref{aM}) where the optomechanical term is dropped.

One can also check that, under  condition (\ref{CondF}), Eq.(\ref{INOUT2}) yields the standard Langevin-framework relation for the output field for the purely optical system, c.f. Eq.(\ref{INOUT}) where the optomechanical term is dropped.

Thus, one can conclude that inequalities (\ref{CondF}) and (\ref{CondF1}) specify the range of applicability of the L-framework when it is  used for the description of the  optical Fabry-Perot cavity.

To conclude this subsection, substitution rules (\ref{rule2}) and (\ref{rule2}) are worth commenting. 
These imply that the input and output fields may be different from those real by phase factors, which cannot be eliminated by the redefining of the phase of the $\textbf{a}$ operator. 
This is in the contrast to the common treatment of $\textbf{A}_{\textrm{in}}$ and $\textbf{A}_{\textrm{out}}$ as the real fields \cite{Walls2008}.
\section{Optomechanical Fabry-Perot cavity: input-output relations approach}
\label{OM}
\subsection{Movable input mirror}
\label{OM1}
Let the input mirror of the Fabry-Perot cavity serve as a mechanical oscillator such that its time-dependent displacement from its average position be $x$, $x>0$ corresponding to an increase of the cavity length (Fig.\ref{FPfig}).
For the treatment of such a problem on the lines of IO-formalism, all needed calculations can be found in Ref.\citenum{Danilishin2012}.
To take into account the time-dependent mirror displacement, neglecting the relativistic corrections, one characterizes the light-mirror coupling with the following equations
\begin{equation}
\label{FPm1}
G_{1}(t)= \rho U_{1}(t - 2x/c)+\tau  U_2(t)
\end{equation}
\begin{equation}
\label{FPm2}
G_{2}(t)= \tau^* U_{1}(t)-\rho^* U_{2}(t+2x/c).
\end{equation}
where all amplitudes are taken at the average position of the mirror.
In the presence of driving field, the above amplitudes are separated into large constant and small fluctuating parts according to Eqs.(\ref{twoparts}).
For the fluctuating parts, in the Fourier domain  and after the linearization with respect to  $x$ and to the fluctuating parts of the fields, Eqs.(\ref{FPm1}) and (\ref{FPm2}) read

\begin{equation}
\label{FPm3}
g_{1}(\Omega)= \rho u_1(\Omega)+\tau  u_2(\Omega)+2i\rho\frac{\omega_L}{c} U_{10}x(\Omega),
\end{equation}
\begin{equation}
\label{FPm4}
g_{2}(\Omega)=\tau^*  u_1(\Omega)-\rho^* u_2(\Omega)+2i\rho^*\frac{\omega_L}{c} U_{20}x(\Omega).
\end{equation}
Since the field amplitudes used are taken at the average (independent of $x$) position of the input mirror, the condition (\ref{ReflCond}) can be used as is such that, with the aid of (\ref{FPMatrix}) and (\ref{link}), Eqs.(\ref{FPm3}) and (\ref{FPm4}) can be rewritten as follows
\begin{equation}
\label{FP31}
u_{1}(\Omega)(e^{-2iL(\Omega+\Delta)/c}-|\rho|)= |\tau| e^{i(\nu-\mu)} u_2(\Omega)+2i|\rho|\frac{\omega_L}{c}U_{10}x(\Omega),
\end{equation}
\begin{equation}
\label{FP41}
e^{i\nu}g_{2}(\Omega)=|\tau| u_1(\Omega)-|\rho| e^{i(\nu-\mu)} u_2(\Omega)+2i|\rho|\frac{\omega_L}{c}U_{20}x(\Omega)e^{i(\nu-\mu)}.
\end{equation}
Now, to  pass  to the quantum equations, which are the most close to those of L-framework, we use the correspondence rules (\ref{rule1})-(\ref{rule11}) appended with the following
\begin{equation}
\label{rule3}
U_{10}\Rightarrow\sqrt{\frac{c}{2L}}a_0,\qquad U_{20}\Rightarrow e^{i(\mu-\nu)}A_0,
\end{equation}
\begin{equation}
\label{rule4}
x(\Omega)\Rightarrow \mathbf{x}(\Omega),
\end{equation}
to find the optomechanical generalizations of (\ref{FP31}) and (\ref{FP41}), which read
\begin{equation}
\label{a1M}
\mathfrak{B}(\Omega) \textbf{a}(\Omega)=\sqrt{\gamma}\textbf{A}_{\textrm{in}}(\Omega)+i|\rho| a_0\frac{\omega_L}{L}\frac{\textbf{x}(\Omega)}{x_{\textrm{zpf}}},
\end{equation}
\begin{equation}
\label{INOUT2M}
|\rho|\textbf{A}_{\textrm{in}}(\Omega)+\textbf{A}_{\textrm{out}}(\Omega)=\sqrt{\gamma}\mathbf{a}(\Omega) +2i|\rho|\frac{\omega_L}{c}A_0\frac{\mathbf{x}(\Omega)}{x_{\textrm{zpf}}}.
\end{equation}

Taking into account that for the system addressed
\begin{equation}
\label{goFP}
g_\omega=\frac{\omega_c}{L} x_{\textrm{zpf}}
\end{equation}
and dropping the terms, which are obviously small to within the accuracy controlled by conditions (\ref{CondF}) and (\ref{CondF1}), one can rewrite Eqs.(\ref{a1M}) and (\ref{INOUT2M}) as follows
\begin{equation}
\label{aFP}
\left[\frac{\gamma}{2}-i(\Omega+\Delta)\right] \textbf{a}(\Omega)
 =\sqrt{\gamma}\textbf{A}_{\textrm{in}}(\Omega)+
 ig_\omega a_0 \frac{\textbf{x}(\Omega)}{x_{\textrm{zpf}}},
\end{equation}
\begin{equation}
\label{INoutFP}
\textbf{A}_{\textrm{in}}(\Omega)+\textbf{A}_{\textrm{out}}(\Omega)=\sqrt{\gamma}\mathbf{a}(\Omega)+i\frac{2L}{c}g_\omega A_0\frac{\textbf{x}(\Omega)}{x_{\textrm{zpf}}}.
\end{equation}

The IO-formalism~\cite{Danilishin2012} also enables the straightforward calculations of operator of the backaction force acting on the input mirror.
Following Ref.\citenum{Tarabrin2013}, in terms of classical physics, the  fluctuating part back-action force reads
\begin{equation}
\label{fRad1}
F(t)= \frac{\hbar \omega_L}{c}[U_{10}^*u_1+ G_{10}^*g_1- U_{20}^*u_2-G_{20}^*g_2]+\textrm{ c.c.}
\end{equation}
For the mirror with the scattering matrix given by Eq.(\ref{FPMatrix}), Eq.(\ref{fRad1}) can be rewritten as
\begin{equation}
\label{fRad2}
  \begin{array}{cc}
F(t)= \frac{ \hbar\omega_L}{c}\left[ \rho(U_{20}^*g_2+G_{10}^*u_1)+\rho^*(G_{20}^*u_2+U_{10}^*g_1) \right] + \textrm{c.c.}= \\
 2\frac{ \hbar\omega_L}{c}\left[ |\rho|^2(U_{10}^*u_1-U_{20}^*u_2)+|\tau||\rho|(e^{i(\mu-\nu)}U_{20}^*u_1+e^{-i(\mu-\nu)}U_{10}^*u_2) \right] + \textrm{c.c.} \\
  \end{array}
\end{equation}
The Fourier transform $\textbf{F}(\Omega)$ of operator of the backaction force, corresponding to the classical force $F(t)$ can be found using the correspondence rules (\ref{rule1})-(\ref{rule11}) and (\ref{rule3}). It reads:
\begin{equation}
\label{fRad3}
\textbf{F}(\Omega) =\frac{\hbar\omega_L}{L} |\rho|^2a_0^*\textbf{a}(\Omega)+ \frac{\hbar\omega_L}{L} |\tau|^2|\rho|\frac{A_0^*\textbf{a}(\Omega)+a_0^*\textbf{A}_{\textrm{in}}(\Omega)}{\sqrt{\gamma}}-\frac{\hbar\omega_L}{L} |\rho|^2A_0^*\textbf{A}_{\textrm{in}}(\Omega)\frac{2L}{c}+
\tilde{\textrm{h}}.\textrm{c}.
\end{equation}
To within the accuracy controlled by conditions (\ref{CondF}) and (\ref{CondF1}), taking into account (\ref{gamma}), Eq.(\ref{fRad3}) implies
\begin{equation}
\label{fRad3a}
\textbf{F}(\Omega)=\frac{\hbar g_\omega}{x_{\textrm{zpf}}}a_0^*\textbf{a}(\Omega)+ \tilde{\textrm{h}}.\textrm{c}.
\end{equation}
In the case where the input noise is white such that $\textbf{A}_{\textrm{in}}(\Omega)$ obeys condition (\ref{comA}) and (\ref{avA}), via (\ref{fRad3a}) and (\ref{aFP}) one finds the following spectral power density of the stochastic back-action force
\begin{equation}
\label{SW1}
S(\Omega)=\gamma\left(
\frac{\hbar |a_0|}{x_{\textrm{zpf}}}\right)^2 
\frac{g_\omega^2}
{(\Delta+\Omega)^2+\gamma^2/4 }.
\end{equation}

\subsection{Movable backstop mirror}
\label{OM2}
Once the backstop mirror serves as the mechanical oscillator, the governing equations for the wave amplitudes read
\begin{equation}
\label{FPm1a}
G_{1}(t)= \rho U_{1}(t)+\tau  U_2(t),
\end{equation}
\begin{equation}
\label{FPm2a}
G_{2}(t)= \tau^* U_{1}(t)-\rho^* U_{2}(t),
\end{equation}
\begin{equation}
\label{ReflCond1}
G_{1} =   - U_{1} e^{-2ik(L+x)}.
\end{equation}
leading to the following equation for the fluctuating parts
\begin{equation}
\label{FPm3a}
g_{1}(\Omega)= \rho u_1(\Omega)+\tau  u_2(\Omega)+2i\rho\frac{\omega_L}{c} U_{10}x(\Omega),
\end{equation}
\begin{equation}
\label{FPm4a}
g_{2}(\Omega)=\tau^*  u_1(\Omega)-\rho^* u_2(\Omega).
\end{equation}
Since (\ref{FPm3a}) is identical to (\ref{FPm3}) one concludes that the derived above Langevin equation (\ref{aFP}) is the same for the cases of movable input and backstop mirrors.
However, in  (\ref{FPm4a}), compared to (\ref{INoutFP}), the last term is missing implying that now
\begin{equation}
\label{IOm}
\textbf{A}_{\textrm{in}}(\Omega)+\textbf{A}_{\textrm{out}}(\Omega)=\sqrt{\gamma}\mathbf{a}(\Omega).
\end{equation}
Thus it turns out, that, for the case the movable backstop mirror, the relation between the fields at the input mirror  is not the same as for the  case of the movable input mirror. 
At the same time, one can check that, to within the accuracy controlled by conditions (\ref{CondF}) and (\ref{CondF1}), the expression (\ref{SW1}) for the spectral power density of the stochastic backaction force is the same in both cases.

\section{Michelson-Sagnac interferometer:input-output relations approach}
\label{Mi}
\begin{figure}
\includegraphics [width=0.7\columnwidth,clip=true, trim=0mm 0mm 0mm 0mm] {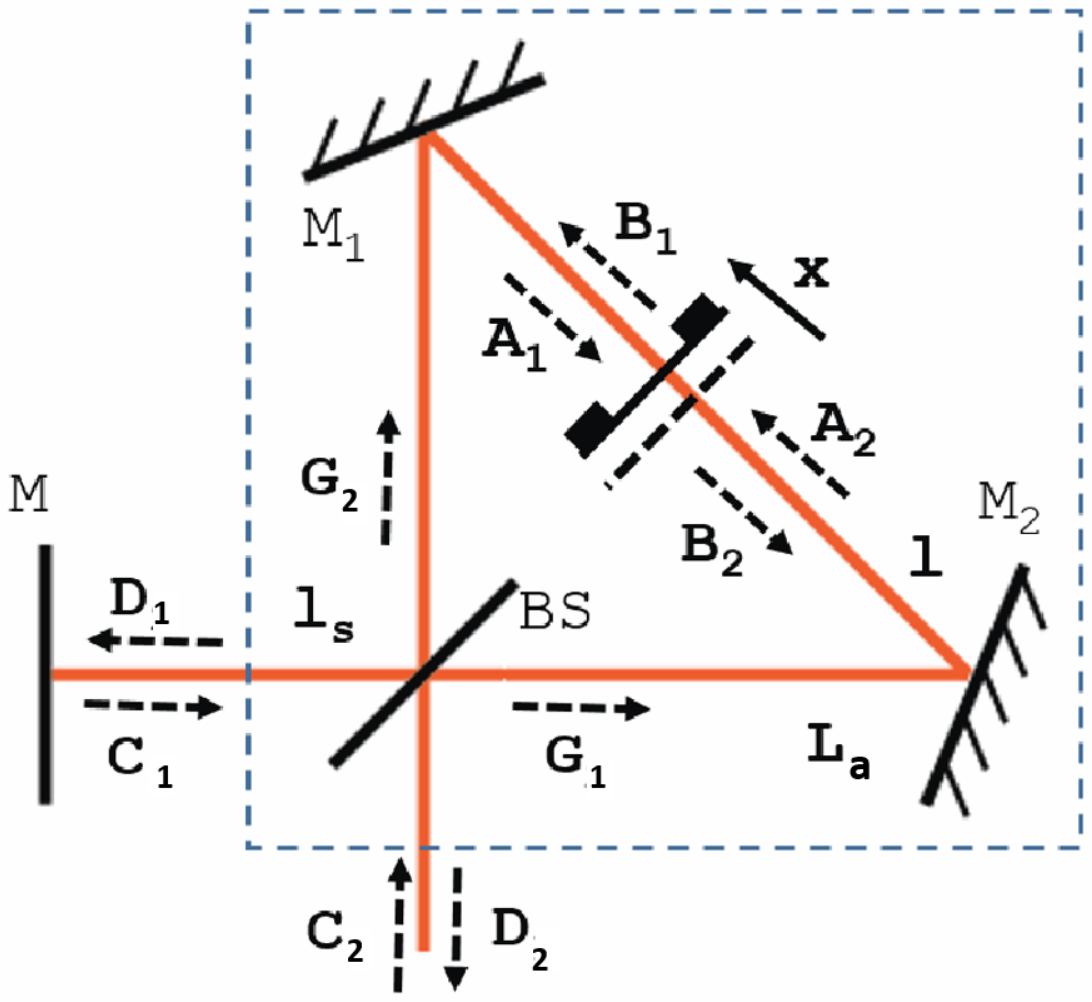}
\caption{Schematics of the modified Michelson-Sagnac interferometer.
The running electromagnetic waves are schematically shown with the arrows and labeled  with their complex amplitudes.
\label{MS}}
\end{figure}
The modified Michelson-Sagnac interferometer (MMSI)~\cite{Tarabrin2013} is schematically depicted  in Fig.\ref{MS}. 
In this setup, the beam splitter (BS) and the membrane provide the following transformations
\begin{equation}
\label{FPMatrixBS}
\left(
  \begin{array}{cc}
G_1 \\
G_2  \\
  \end{array}
\right)= \left(
  \begin{array}{cc}
T &  -R \\
R  &  T  \\
  \end{array}
\right)
 \left(
  \begin{array}{cc}
C_1 \\
C_2  \\
  \end{array}
\right)
\qquad \textrm{and}
\qquad
\left(
  \begin{array}{cc}
B_1 \\
B_2  \\
  \end{array}
\right)= \left(
  \begin{array}{cc}
-r &  t \\
t  &  r \\
  \end{array}
\right)
 \left(
  \begin{array}{cc}
A_1 \\
A_2  \\
  \end{array}
\right)
\end{equation}
of amplitudes of the waves indicated in this figure, where all coefficients of the matrices are real and positive.
Mirrors $M$, $M_1$, and  $M_2$ are ideal, imposing the $\pi$ phase shift on reflection.
The membrane is displaced from its symmetric position (shown with the dashed line in the figure) by distance $x$ in the direction of the arrow.
The BS-M1 and BS-M2 distances are equal to $L_a$.
The M1-M2 distance equals $2l$.
The M-BS distance equals $l_s$.
The part of MMSI marked with the dashed rectangle can be considered as an effective mirror, which is characterized by a scattering matrix~\cite{Xuereb2011,Tarabrin2013} depending on $x$.
Specifically, the amplitudes of the wave amplitudes $D_1$, $D_2$, $C_1$, and $C_2$ (Fig.~\ref{MS}) are linked by the following relation~\cite{Tarabrin2013}
\begin{equation}
\label{MSlink}
\left(
  \begin{array}{cc}
D_1 \\
D_2  \\
  \end{array}
\right)=e^{2ik(L_a+l)}\mathbb{M}
 \left(
  \begin{array}{cc}
C_1 \\
C_2  \\
  \end{array}
\right),
\end{equation}
\begin{equation}
\label{MSMatrix}
\mathbb{M}= \left(
  \begin{array}{cc}
\rho & \tau \\
\tau &-\rho^* \\
  \end{array}
\right), \qquad \rho=|\rho| e^{i\mu}, \qquad \qquad \tau=|\tau| e^{i\nu},
\end{equation}
\begin{equation}
\label{MSrho}
\rho=  -\beta t +\alpha r \cos\psi + ir \sin\psi,
\end{equation}
\begin{equation}
\label{MStau}
\tau= t\alpha+\beta r \cos\psi.
\end{equation}
\begin{equation}
\label{def}
\psi= 2ck, \qquad \alpha =T^2-R^2, \qquad \beta =2RT, \qquad \alpha^2+\beta^2=1.
\end{equation}
The interferometer decay rate associated with this effective mirror is~\cite{Tarabrin2013}
\begin{equation}
\label{MSgamma}
\gamma= \frac{\tau^2c}{2 L}
\end{equation}
where $L = L_a+l+l_s$.
It was shown~\cite{Xuereb2011,Tarabrin2013} that the system behaves as a one-sided Fabry-Perot cavity with the optical length $L$, the input mirror of which  is characterized by the scattering matrix $\mathbb{M}$ such that the cavity resonance frequencies read
\begin{equation}
\label{omega}
\omega_c= \frac{c}{2L}(2\pi N -\mu)
\end{equation}
where $N$ is integer.

Thus, one can define the optomechanical coupling constants:
\begin{equation}
\label{go3}
g_\omega=-\frac{d\omega_c}{dx} x_{\textrm{zpf}}=g_0\frac{d\mu}{d\psi},
\qquad g_\gamma=-\frac{1}{2}\frac{d\gamma}{dx}=-g_0\tau\frac{d\tau}{d\psi}
\qquad g_0= \frac{ ck}{L}x_{\textrm{zpf}}
\end{equation}
with
\begin{equation}
\label{derivatives}
\frac{d\mu}{d\psi} = r\frac{\alpha r- \beta t \cos\psi }{(\beta t - \alpha r\cos\psi)^2 + r^2\sin\psi^2 }, \qquad
\tau\frac{d\tau}{d\psi} =-\beta r\sin\psi (t\alpha+\beta r \cos\psi).
\end{equation}

In IO-formalism, as a starting point one uses the input-output relations at the effective input mirror where the coefficients of the scattering matrix are $x$-dependent to write
\begin{equation}
\label{SMm1}
D_{1}(t)= \rho(x) C_{1}(t)+\tau(x)  C_2(t)
\end{equation}
\begin{equation}
\label{SMm2}
D_{2}(t)= \tau (x) C_{1}(t)-\rho(x)^* C_{2}(t).
\end{equation}

 Let the system be excited with a strong monochromatic light of frequency  $\omega_\textrm{L}$.
 We are interested in the impact of small, compared to the light wavelength, time-dependent displacements of the membrane about its average position.
We keep $x$ as a notation for the average position while we denote the Fourier transform of the small displacements as $x(\Omega)$.
In the presence of the driving field, we consider the above amplitudes in the reference frame rotating with frequency $\omega_\textrm{L}$ separating the large constant and small fluctuating parts like in Eq.(\ref{twoparts}).
In the Fourier domain, we linearize the problem with respect to $x(\Omega)$ and to the  fluctuating parts of the fields.
Thus, Eqs.(\ref{SMm1}) and (\ref{SMm2}) imply
\begin{equation}
\label{SM11}
d_{1}(\Omega)=\tau c_{2}(\Omega)+ \rho c_{1}(\Omega) +\frac{\partial\tau}{\partial x} C_{20}x(\Omega)+\frac{\partial\rho}{\partial x} C_{10}x(\Omega),
\end{equation}
\begin{equation}
\label{SM22}
d_{2}(\Omega)=-\rho^* c_{2}
(\Omega)+\tau c_{1}(\Omega) -\frac{\partial\rho^*}{\partial x}
C_{20}x(\Omega)
 + \frac{\partial\tau}{\partial x} C_{10}x(\Omega)
\end{equation}
where, using (\ref{go3}) and (\ref{derivatives}), 
\begin{equation}
\label{rodx}
\frac{\partial\rho}{\partial x}
= \frac{2L}{c}
\frac{e^{i\mu}}{x_{\mathrm{zpf}}}
(i|\rho|g_\omega+ g_\gamma/|\rho|),
\end{equation}
\begin{equation}
\label{tadx}
\frac{\partial\tau}{\partial x}
= -\sqrt{\frac{2L}{c}}
\frac{e^{i\nu}}{x_{\mathrm{zpf}}}\frac{ g_\gamma}{\sqrt{\gamma}}.
\end{equation}
Note that when writing (\ref{SM11}) and (\ref{SM22}) we neglected the $\Omega$-dependence of $\tau$ and $\rho$.
In view of the goal of the comparison with the results of the Langevin-equation formalism, such a neglect is acceptable.

On the lines of Sect.\ref{O},  using  (\ref{rodx}), (\ref{tadx}) and the following correspondence rules
\begin{equation}
\label{rulea}
c_{1}(\Omega)\Rightarrow\sqrt{\frac{c}{2L}}\textbf{a}(\Omega),\qquad c_{2}(\Omega)\Rightarrow  e^{i(\mu-\nu)}\textbf{A}_{\textrm{in}}(\Omega),
\end{equation}

\begin{equation}
\label{ruleb}
d_{2}(\Omega)\Rightarrow e^{-i\nu}\textbf{A}_{\textrm{out}}(\Omega),\qquad x(\Omega)\Rightarrow \mathbf{x}(\Omega),
\end{equation}
\begin{equation}
\label{rulec}
C_{10}\Rightarrow\sqrt{\frac{c}{2L}}a_0,\qquad C_{20}\Rightarrow e^{i(\mu-\nu)}A_0,
\end{equation}
Eqs. (\ref{SM11}) and (\ref{SM22}) appended with  the reflection condition at the backstop mirror $M$ yield
\begin{equation}
\label{aMQ}
\mathfrak{B}(\Omega)\textbf{a}(\Omega)
 =\sqrt{\gamma}\textbf{A}_{\textrm{in}}(\Omega)+\frac{\textbf{x}(\Omega)}{x_{\textrm{zpf}}}
\left[ ig_\omega|\rho| a_0 + g_\gamma(a_0/|\rho|- A_0/\sqrt{\gamma}) \right ],
\end{equation}
\begin{equation}
\label{INOUT3}
\textbf{A}_{\textrm{out}}(\Omega)+|\rho|\textbf{A}_{\textrm{in}}(\Omega)=\sqrt{\gamma}\mathbf{a}(\Omega) +i\frac{2 L}{c}g_\omega A_0
\textbf{x}(\Omega)
-\frac{g_\gamma a_0}{\sqrt{\gamma}}\textbf{x}(\Omega)
-\frac{2L}{c}\frac{ g_\gamma A_0}{|\rho|}
\textbf{x}(\Omega).
\end{equation}
Dropping the terms, which are obviously small to within the accuracy controlled by conditions (\ref{CondF}) and (\ref{CondF1}), these equations can be rewritten as follows %
\begin{equation}
\label{aMM}
\left[\frac{\gamma}{2}-i(\Omega+\Delta)\right] \textbf{a}(\Omega)
 =\sqrt{\gamma}\textbf{A}_{\textrm{in}}(\Omega)+
\left[ ig_\omega a_0 + g_\gamma(a_0- A_0/\sqrt{\gamma}) \right ]\frac{\textbf{x}(\Omega)}{x_{\textrm{zpf}}},
\end{equation}
\begin{equation}
\label{INOUT4}
\textbf{A}_{\textrm{out}}(\Omega)+\textbf{A}_{\textrm{in}}(\Omega)=\sqrt{\gamma}\mathbf{a}(\Omega) +i\frac{2 L}{c}g_\omega A_0
\frac{\textbf{x}(\Omega)}{x_{\textrm{zpf}}}
-\frac{g_\gamma a_0}{\sqrt{\gamma}} \frac{\textbf{x}(\Omega)}{x_{\textrm{zpf}}}.
\end{equation}

For MMSI, the calculations of the stochastic backaction force via the momentum-balance equation is straightforward but cumbersome~\cite{Tarabrin2013}.
These calculations are outlined in Appendix \ref{Pres}.
Here we will give the result for the Fourier transform $\textbf{F}(\Omega)$ of the backaction force operator where the terms which are small in view of  inequalities (\ref{CondF}) and (\ref{CondF1}) are dropped: 
\begin{equation}
\label{F3}
\begin{array}{cc}
\textbf{F}(\Omega)=\frac{\hbar g_\omega}{x_{\textrm{zpf}}}a_0^*\textbf{a}(\Omega)
+i\frac{\hbar g_\gamma}{x_{\textrm{zpf}}}\frac{a_0^*\textbf{A}_{\textrm{in}}(\Omega)-A_0^*\textbf{a}(\Omega)}{\sqrt{\gamma}}
+ \tilde{\textrm{h}}.\textrm{c}.
  \end{array}
\end{equation}
Using (\ref{F3}) we arrive to the following expression
\begin{equation}
\label{SWM}
S(\Omega)=\gamma\left(
\frac{\hbar | a_0|}{x_{\textrm{zpf}}}\right)^2
\frac{g_\omega^2}
{(\Delta+\Omega)^2+\gamma^2/4 }\left(1 + \frac{g_\gamma}{g_\omega}\frac{2\Delta+\Omega}{\gamma}\right) ^2.
\end{equation}
for the spectral power density of the stochastic backaction force. 

\section{Discussion}
\label{Di}
\subsection{The generation of the dissipative coupling: the Langevin equation and the spectral power density of the stochastic backaction force}
\label{Generation}
Let us compare the results for the field in the cavity and the spectral power density of the stochastic backaction force obtained for using  IO-formalism with those obtained using L-formalism.

For the optomechanical Fabry-Perot cavity, we compare (\ref{aFP}) and (\ref{SW1}) with (\ref{aM}) and (\ref{S}), respectively.
It is seen that both approaches demonstrate the presence of the dispersive coupling whereas the presence of the dissipative coupling predicted in this system by L-formalism is not justified by IO-formalism. 

In contrast, for MMSI, as clear from the results of IO-formalism, in Eqs. (\ref{aMM}) and (\ref{SWM}), the dissipative coupling is present  justifying the result of L-formalism, Eqs. (\ref{aM}) and (\ref{S}).
In Ref.\citenum{Xuereb2011}, MMSI was identified as a system where the regime with the purely dissipative coupling is possible and is of interest. 
In this context, it is worth evaluating the the strength of this coupling in this regime.
Using (\ref{go3}) and (\ref{derivatives}), one finds that the regime where the coupling is purely dissipative acarus at $\cos\psi= \alpha r/(\beta t)$ on the condition that $|\alpha r/(\beta t)| \leq 1$, implying that
\begin{equation}
\label{tauF}
|g_\gamma|= g_0\beta r\sqrt{1 -a^2r^2(\beta^2 t^2)}|\tau| < g_0|\tau| . 
\end{equation}
Since $|\tau| \ll 1$,  $|g_\gamma|$ is small  compared to $g_0$, which is  the value of the dispersive coupling constant in the optomechanical cavity having the length $L$.
Note that much higher values of $|g_\gamma|/g_0$ were predicted for alternative systems with the purely dissipative optomechanical coupling \cite{tagantsev2020a,tagantsev2021}   

The above difference between the Fabry-Perot and the MMSI system can be rationalized as follows. 
For a general optomechanical one-sided cavity with the input mirror  power transmission $|\tau|^2$ and the length $L$ (for MMSI these are effective), we can write
\begin{equation}
\label{ggg}
g_\gamma=-\frac{x_\textrm{zpf}}{2}\frac{d\gamma}{dx} =\frac{x_\textrm{zpf}}{2}\left[|\tau|^2 \frac{c}{2L^2}\frac{dL}{dx}-\frac{d|\tau|^2}{dx}\frac{c}{2L}\right].
\end{equation}
In the Fabry-Perot system,  only the first term in the bracket in (\ref{ggg}) is nonzero while, in the MMSI system, it is the second one.
In other words, the dissipative coupling is generated by the dependence of the photon escape probability on the oscillator coordinate while the coordinate dependence of the cavity decay rate may not lead to such a generation.
Note that this statement should hold not only for MMSI but also for any system that can be presented as an optical resonator with  photon escape probability dependent on the oscillator coordinate. 
\subsection{The boundary condition at the input mirror}
\label{additionalB}
Let us start with the Fabry-Perot system. First of all we note that, in terms of L-formalism, the boundary condition at the input mirror Eq.(\ref{INOUT}) contains the term that  is supposed to be generated by dissipative coupling. 
IO-formalism, as seen from Eq.(\ref{INoutFP}), does not confirm such a generation.
This is consistent with aforementioned absence of this coupling in the system.

For the Fabry-Perot system, there exists another difference between the boundary condition generated by L-formalism and that generated by IO-formalism. 
In L-formalism, for the two situations,  where mechanically active input is mechanically active and where the backstop  mirror is, the boundary conditions are the same. 
At the same time, according to IO-formalism, this is not the case, c.f. (\ref{INoutFP}) with  (\ref{IOm}).
It is seen that,  in the case of the mechanically active backstop mirror, the result obtained using L-formalism is justified while, if it is the input mirror which is  mechanically active, the following term 
\begin{equation}
\label{add1}
 \textbf{A}_{x1}(\Omega)=i\frac{2L}{c}g_\omega A_0\frac{\textbf{x}(\Omega)}{x_{\textrm{zpf}}}.
\end{equation}
should be incorporated in the boundary condition.
The physical meaning of this term is transparent: it is the modulation of the light reflected from the moving input mirror. 
Note that, as it should be, this term is independent of $L$.

To assess the importance of this term, one should compare it with  the $\mathbf{x}$-dependent contribution to $\textbf{A}_{\textrm{out}}$, which is due to the $\mathbf{x}$-dependent part of $\sqrt{\gamma}\mathbf{a}(\Omega)$.
Via (\ref{aFP}) and (\ref{INoutFP}), this contribution reads
\begin{equation}
\label{ax}
 \textbf{A}_{x2}(\Omega)=\frac{i\sqrt{\gamma}a_0}{\gamma/2 -i(\Delta+\Omega)}\frac{ g_\omega}{x_{\textrm{zpf}}}\textbf{x}(\Omega).
\end{equation}
The ratio between (\ref{add1}) and (\ref{ax}), $R$,  can be evaluated as follows
\begin{equation}
\label{ratio}
R= \frac{ \textbf{A}_{x1}}{ \textbf{A}_{x2}}
= \frac{ 2\pi}{ \gamma \omega_{\textrm{FSR}}}
[\gamma/2-i(\Delta+\Omega)][\gamma/2-i\Delta].
\end{equation}
If  $|R|\ll1$, the term given by Eq.(\ref{add1}) can be neglected in the boundary condition. 
Considering the expression for $|R|^2$:
\begin{equation}
\label{ratioSQ}
|R|^2= 
= \frac{ 4\pi^2}{ \gamma^2 \omega_{\textrm{FSR}}^2}
[\gamma^2/4+(\Delta+\Omega)^2][\gamma^2/4+\Delta^2],
\end{equation}
one readily finds that this condition requires that
\begin{equation}
\label{condSUP}
\gamma \gg \frac{2\pi|\Delta(\Delta+\Omega)|}{\omega_{\textrm{FSR}}}. 
\end{equation}
Taking into account that, staying in range of applicability of L-formalism where
\begin{equation}
\label{CondF11}
\gamma \gg \frac{2\pi(\Delta+\Omega)^2}{\omega_{\textrm{FSR}}}, 
\end{equation}
we conclude that, at
\begin{equation}
\label{CondF111}
|\Delta+\Omega|>|\Delta|,
\end{equation}
$|R|\ll1$ such that the relation
\begin{equation}
\label{INOUT0}
\textbf{A}_{\textrm{in}}(\Omega)+\textbf{A}_{\textrm{out}}(\Omega)=\sqrt{\gamma}\mathbf{a}(\Omega),
\end{equation}
which is commonly used in L-formalism in the absence of the dissipative coupling,  is applicable.
At the sane time, at
\begin{equation}
\label{CondF112}
|\Delta+\Omega|\ll|\Delta|,
\end{equation}
there appears a regime where
\begin{equation}
\label{CondF12}
 \frac{2\pi|\Delta(\Delta+\Omega)|}{\omega_{\textrm{FSR}}}\geq\gamma \gg \frac{2\pi(\Delta+\Omega)^2}{\omega_{\textrm{FSR}}}. 
\end{equation}
In this regime, $|R|\geq 1$  such that the term given by Eq.(\ref{add1}) must betaken into account and 
the modified relation
\begin{equation}
\label{INoutFP0}
\textbf{A}_{\textrm{in}}(\Omega)+\textbf{A}_{\textrm{out}}(\Omega)=\sqrt{\gamma}\mathbf{a}(\Omega)+i\frac{2L}{c}g_\omega A_0\frac{\textbf{x}(\Omega)}{x_{\textrm{zpf}}}.
\end{equation}
should be used instead of  Eq.(\ref{INOUT0}).

The above discussion from this subsection was dealing with the  Fabry-Perot system.
As for MMSI, the boundary condition at the input mirror, according to Eq.(\ref{INOUT4}), reads
\begin{equation}
\label{INOUTfinal}
\textbf{A}_{\textrm{out}}(\Omega)+\textbf{A}_{\textrm{in}}(\Omega)=\sqrt{\gamma}\mathbf{a}(\Omega) +i\frac{2 L}{c}g_\omega A_0
\frac{\textbf{x}(\Omega)}{x_{\textrm{zpf}}}
-\frac{g_\gamma a_0}{\sqrt{\gamma}} \frac{\textbf{x}(\Omega)}{x_{\textrm{zpf}}}.
\end{equation}

 In this system, since the dissipative coupling is active,
 Eq.(\ref{INOUTfinal}) contains the terms with $g_\gamma$.
Here, the term given by Eq.(\ref{add1}) is also present.
One can readily check that, like for the Fabry-Perot system, for MMSI, this term must be taken into account only in the regime specified by inequalities
(\ref{CondF112}) and (\ref{CondF12}).
\subsection{Phase factor}
\label{Phase}
As is clear from the above considerations  when L-formalism is applied to a purely  optical or optomechanical system, the complex amplitude $A_0$ as well as  operators $\textbf{A}_{\textrm{in}}$ and  $\textbf{A}_{\textrm{out}}$ may not directly correspond to the real fields outside the cavity.
It is seen from  the correspondence rules (\ref{rule2}), (\ref{rule11}), (\ref{rule3}), (\ref{rulea}), (\ref{ruleb}), and (\ref{rulec}), that there may be a phase shift difference, i.e. a difference in a phase factor like $e^{i\varphi}$.
Note that we introduced the ladder operator for the cavity field without any additional phase factor.
One can readily check that the possible existence of the above phase factor  could be recognized already at the initial phenomenological formulation of the L-formalism~\cite{Collet1984}.
Specifically, in Ref.~\cite{Collet1984} in Eq.(2), the factor $\gamma'$, which links the input term with the operator of the input field, was assumed to be real.
If such an assumption were lifted, the consideration of this paper would yield an equation for the $\textbf{a}$-operator where the input field enters with an unknown phase factor that cannot be specified in terms of the phenomenological framework.
As for the derivation~\cite{Walls2008} based on Hamiltonian (\ref{Ham}), it presents no demonstration that  operators $\textbf{A}_{\textrm{in}}$ and  $\textbf{A}_{\textrm{out}}$ do correspond to the real fields outside the cavity.
In the standard practice, one assumes an exact correspondence between the operators for the fields outside the cavity and the physical fields outside it.

It this context, one should mention that, for the 40-years of use of the L-formalism, the aforementioned  phase factor being neglected, no problem with an inconsistent description was reported.
The reason for that is that the systems addressed were mainly purely optical or optomechanical governed by the dispersive coupling.
One readily checks that, in this case,  the phase difference between the field inside the cavity and outside it never enter the game such that the results should be insensitive to the phase factor in question.
At the same time, it is clear from the form of the dissipative coupling  terms containing $g_\gamma$ in Eq.(\ref{aMM})
\begin{equation}
\label{term}
g_\gamma(a_0- A_0/\sqrt{\gamma})\frac{\textbf{x}(\Omega)}{x_{\textrm{zpf}}}
\end{equation}
that the description of this coupling is really sensitive to the above phase factors.
The point is that, in terms of the standard L-formalism, 
 the  large  amplitudes $a_0$ and $A_0$, to within the factor of $\sqrt{\frac{c}{2L}}$ are equal to the  amplitudes entering the equation of the classical optics, i.e. $C_{10}(\Omega)$ and $C_{20}$ in our notation.
 However, in view of the corresponding rules (\ref{rulec}):
$$
C_{10}\Rightarrow\sqrt{\frac{c}{2L}}a_0,\qquad C_{20}\Rightarrow e^{i(\mu-\nu)}A_0
$$
it is clear that the use of non-phase corrected values of classical amplitudes when writing (\ref{term}) may lead to essentially incorrect values of this term.

The aforementioned phase factor must also to be taken into account when quantizing the problem which involves the dissipative coupling. 
The quantization of MMSI, which does not take into account the phase factors, was done in Ref.~\cite{Xuereb2011}.
In this paper, the effective transparency of input mirror and the dissipative coupling constant as functions of the mechanical oscillator coordinate ($\tau $, $g_\gamma$ and $x$  in our notation, respectively) were evaluated.
Using those functions, one readily checks that  $g_\gamma$  does not scale as $\frac{d|\tau|^2 }{dx}$.
This is in conflict with the result obtained in Sec.\ref{Mi}.
We believe that this disparity is the results of neglect of the aforementioned phase factor.
\section{Summary}
\label{Co}
The analysis of an optomechanical Fabry-Perot  cavity and a modified Michelson-Sagnac interferometer (MMSI) by using a model-free input-output relations approach~\cite{Buonanno2003,Danilishin2012,Khalili2016} (IO-formalism) was used to derive the relations of quantum  Langevin equation formalism (L-formalism).
This enables us to critically review the descriptive capability of the latter.

It was shown that the validity of the one-mode Langevin equation  for the intracavity in an optical or optomechanical cavity  requires that
\begin{equation}
\label{CondTot}
 \frac{\omega_{\textrm{FSR}}}{2\pi}
\gg \gamma\gg \frac{2\pi(\Delta+\Omega)^2}{\omega_{\textrm{FSR}}}.
\end{equation}
This condition is more stringent than the  intuitive condition $\gamma,\Omega,\Delta\ll\omega_{\textrm{FSR}}$. 
Remarkably, (\ref{CondTot}) sets the low limit for the cavity decay rate. 

It was demonstrated that the dependence of the cavity decay rate on the coordinate of the mechanical oscillator does not necessarily lead to the generation of the dissipative optomechanical coupling  in addition to that dispersive.
This is in contrast to the standard L-formalism \cite{Elste2009}.
We found that in the optomechanical Fabry-Perot cavity it is not generated while  in the case of MMSI it is.
From this fact, it was concluded that the generation of dissipative coupling requires the dependence of  transparency of the input mirror (real or effective) on the coordinate of the mechanical oscillator.

It was shown  that, when the optomechanical Fabry-Perot cavity is used to get the information about mechanics,  the relation between  the incident and reflected fields at the input mirror may not be the same for the cases where the input mirror plays the role of the mechanical oscillator and where  the backstop mirror does.
This relation is the same, having its standard form,  only if
\begin{equation}
\label{CondF113}
|\Delta+\Omega|>|\Delta|.
\end{equation}
Otherwise, for the case where the input mirror plays the role of the mechanical oscillator, there exists a regime where it should be modified.
For MMSI, in such a regime, the same modification of also needed,
The L-formalism does not distinguish those situations.

It was  found that, the in L-formalism, in general, the intracavity field and the field outside the cavity cannot be at the same time associated with the real fields.
In reality this can be done up to a phase factor.
Commonly, this factor was neglected in calculations, typically creating no problem.
The reason for that is that the systems addressed were mainly purely optical or those optomechanical controlled by the dispersive coupling.
However, if the dissipative coupling is involved, the phase factor may  matter.
In general, it should be taken into account at quantization of optomechanical problems.
\section{Acknowledgements}
 Prof. I.V.Sokolov is acknowledged for fruitful discussions. 
\appendix
\section{Backaction in modified Michelson-Sagnac interferometer}
\label{Pres}
The expression for the backaction force in MMSI can be derived from a more general results obtained by Tarabrin et al \cite{Tarabrin2013} for the so-called signal-recycled Michelson-Sagnac interferometer.
Here we outline the calculations for the back-action force simplified down to the accuracy of the Langevin-equation formalism.

On the lines of IO-formalism, for MMSI, the calculations of the operator of  back-action force in terms of the operators $\textbf{A}_{\textrm{in}}(\Omega)$ and $\textbf{a}(\Omega)$ are more cumbersome  than those in the case of the Fabry-Perot cavity but  similar.
In terms of the complex  amplitudes of the fields interacting with the membrane $A_1$, $A_2$, $B_1$, and $B_2$ (Fig.\ref{MS}) and the parameters of its scattering matrix, Eq.(\ref{FPMatrixBS}), the Fourier transform of Eq.(\ref{fRad2}) can be rewritten as follows
\begin{equation}
\label{fRad4}
  \begin{array}{cc}
<F(\Omega)>=
 2r\frac{ \hbar\omega_L}{c}\left[ r(A_{20}^*a_2(\Omega)-A_{10}^*a_1(\Omega))+t(A_{10}^*a_2(\Omega)+A_{20}^*a_1(\Omega)) \right] + \textrm{c.c.} \\
  \end{array}
\end{equation}
Hereafter, similar to Eq.(\ref{twoparts}), we use the upper and lower case symbols to denote the constant and fluctuating parts of the fields, respectively.

Next, we pass to the amplitudes $C_1$, and $C_2$ (Fig.\ref{MS}) using the following evident relations
\begin{equation}
\label{transform}
\left(
  \begin{array}{cc}
a_1(\Omega) \\
a_2 (\Omega) \\
  \end{array}
\right)= \mathbb{T}(\Omega)
 \left(
  \begin{array}{cc}
c_1 (\Omega)\\
c_2 (\Omega) \\
  \end{array}
\right)
\qquad \textrm{and}
\qquad
\left(
  \begin{array}{cc}
A_{10} \\
A_{20} \\
  \end{array}
\right)= \mathbb{T}(0)
 \left(
  \begin{array}{cc}
C_{10} \\
C_{20} \\
  \end{array}
\right)
\end{equation}
\begin{equation}
\label{T}
\mathbb{T}(\Omega)= e^{ik(L_a+l)}\left(
  \begin{array}{cc}
-Re^{-ikx} &  Te^{-ikx} \\
Te^{ikx} & Re^{ikx} \\
  \end{array}
\right)
\qquad \mathrm{where}\qquad k=(\omega_L +\Omega)/c,
\end{equation}
to find
\begin{equation}
\label{fRad5}
<F(\Omega)>=
2r\frac{ \hbar\omega_L}{c} e^{i\frac{\Omega}{c}(L_a+l)}
\left\{ P[(C_{10}^*c_1(\Omega)-C_{20}^*c_2(\Omega)]+ QC_{20}^*c_1(\Omega) + Q^*C_{10}^*c_2(\Omega) \right\} + \textrm{c.c.},
\end{equation}
\begin{equation}
\label{PQ}
P=r(T^2-R^2) -2tTR\cos2kx\qquad Q=t(T^2e^{2ikx}-R^2e^{-2ikx})+2rTR.
\end{equation}
To translate Eq.(\ref{fRad5}) into quantum mechanics, we use the following correspondence rules
\begin{equation}
\label{rule5}
  \begin{array}{cc}
c_{2}(\Omega)\Rightarrow \textrm{sign}\tau e^{i\mu}\textbf{A}_{\textrm{in}}(\Omega), &   c_{1}(\Omega)\Rightarrow\sqrt{\frac{c}{2L}}\textbf{a}(\Omega) \\
C_{20}\Rightarrow \textrm{sign}\tau e^{i\mu}A_0, &  C_{10}\Rightarrow\sqrt{\frac{c}{2L}}a_0, \\
  \end{array}
\end{equation}
which take into account that the system in question can be viewed as a Fabry-Perrot interferometer with the optical length $L = L_a+l+l_s$ and the scattering matrix of the input mirror given by Eq.(\ref{MSMatrix}).
Thus, via straightforward but cumbersome calculations we arrive at the following expression for the operator of the backaction force
\begin{equation}
\label{fRad6}
\begin{array}{cc}
\textbf{F}(\Omega)=\frac{\hbar g_\omega}{x_{\textrm{zpf}}}a_0^*\textbf{a}(\Omega)
+i\frac{\hbar g_\gamma}{x_{\textrm{zpf}}}\frac{a_0^*\textbf{A}_{\textrm{in}}(\Omega)-A_0^*\textbf{a}(\Omega)}{\sqrt{\gamma}}
+\tau^2\frac{\hbar g_\omega}{x_{\textrm{zpf}}}\frac{a_0^*\textbf{A}_{\textrm{in}}(\Omega)+A_0^*\textbf{a}(\Omega)}{\sqrt{\gamma}} \\
  -\frac{\hbar g_\omega}{x_{\textrm{zpf}}}A_0^*\textbf{A}_{\textrm{in}}(\Omega)\frac{2L}{c}+ \tilde{\textrm{h}}.\textrm{c}., \\
  \end{array}
\end{equation}
\begin{equation}
\label{go4}
\frac{g_\omega}{ x_{\textrm{zpf}}}=\frac{r\omega_L}{L}P=\frac{d\omega_c}{dx},
\end{equation}
\begin{equation}
\label{gg4}
\frac{g_\gamma}{ x_{\textrm{zpf}}}=\frac{r\omega_L}{L}\tau\textrm{Im}[e^{-i\mu}Q]=-\frac{1}{2}\frac{d\gamma}{dx},
\end{equation}
\begin{equation}
\label{gs}
\tau^2\frac{g_\omega}{ x_{\textrm{zpf}}}=\frac{r\omega_L}{L}\tau\textrm{Re}[e^{-i\mu}Q].
\end{equation}
When writing Eqs. (\ref{go4})-(\ref{gs}) the terms, which are evidently small within the approximation controlled  by inequalities (\ref{CondF}), were neglected.
Within the same approximation, in $P$ and $Q$ entering Eqs.(\ref{go4}), (\ref{gg4}), and (\ref{gs}), it is set $k=\omega_c/c$.

Using (\ref{fRad6}) and (\ref{aM}) one finds the following operator of the stochastic backaction force
\begin{equation}
\label{fRadF}
\textbf{F}(\Omega)=\frac{\hbar \sqrt{\gamma}}{x_{\textrm{zpf}}}
\frac{ g_\omega+ g_\gamma \frac{2\Delta+\Omega}{\gamma} -ig_\omega \frac{L\Omega}{c}}{\gamma/2 -i(\Delta+\Omega)}
a_0^*\textbf{A}_{\textrm{in}}+ \tilde{\textrm{h}}.\textrm{c}.
\end{equation}
This equation is consistent with a more general result from Ref.\cite{Tarabrin2013} for the recycled Michelson-Sagnac interferometer, once it is  adjusted to our setting and taken in the approximation we adopt.
It is worth mentioning that, in this equation, the imaginary term in the numerator is  due to the third and the fourth terms in (\ref{fRad6}), which give  $-2ig_\omega \frac{L\Omega}{c}$ and  $ig_\omega \frac{L\Omega}{c}$, respectively.
Equation (\ref{fRadF}) implies the following expression for the spectral power density of the stochastic backaction force
\begin{equation}
\label{SF1}
S(\Omega)=\gamma \left(
\frac{\hbar| a_0|}{x_{\textrm{zpf}}}\right)^2
\frac{\left(g_\omega +g_\gamma\frac{2\Delta+\Omega}{\gamma}\right) ^2+\left(g_\omega \frac{L\Omega}{c}\right)^2}
{(\Delta+\Omega)^2+\gamma^2/4 }.
\end{equation}
\bibliography{FOR3,NF}

\begin{thebibliography}{36}%
\makeatletter
\providecommand \@ifxundefined [1]{%
 \@ifx{#1\undefined}
}%
\providecommand \@ifnum [1]{%
 \ifnum #1\expandafter \@firstoftwo
 \else \expandafter \@secondoftwo
 \fi
}%
\providecommand \@ifx [1]{%
 \ifx #1\expandafter \@firstoftwo
 \else \expandafter \@secondoftwo
 \fi
}%
\providecommand \natexlab [1]{#1}%
\providecommand \enquote  [1]{``#1''}%
\providecommand \bibnamefont  [1]{#1}%
\providecommand \bibfnamefont [1]{#1}%
\providecommand \citenamefont [1]{#1}%
\providecommand \href@noop [0]{\@secondoftwo}%
\providecommand \href [0]{\begingroup \@sanitize@url \@href}%
\providecommand \@href[1]{\@@startlink{#1}\@@href}%
\providecommand \@@href[1]{\endgroup#1\@@endlink}%
\providecommand \@sanitize@url [0]{\catcode `\\12\catcode `\$12\catcode `\&12\catcode `\#12\catcode `\^12\catcode `\_12\catcode `\%12\relax}%
\providecommand \@@startlink[1]{}%
\providecommand \@@endlink[0]{}%
\providecommand \url  [0]{\begingroup\@sanitize@url \@url }%
\providecommand \@url [1]{\endgroup\@href {#1}{\urlprefix }}%
\providecommand \urlprefix  [0]{URL }%
\providecommand \Eprint [0]{\href }%
\providecommand \doibase [0]{http://dx.doi.org/}%
\providecommand \selectlanguage [0]{\@gobble}%
\providecommand \bibinfo  [0]{\@secondoftwo}%
\providecommand \bibfield  [0]{\@secondoftwo}%
\providecommand \translation [1]{[#1]}%
\providecommand \BibitemOpen [0]{}%
\providecommand \bibitemStop [0]{}%
\providecommand \bibitemNoStop [0]{.\EOS\space}%
\providecommand \EOS [0]{\spacefactor3000\relax}%
\providecommand \BibitemShut  [1]{\csname bibitem#1\endcsname}%
\let\auto@bib@innerbib\@empty
\bibitem [{\citenamefont {Aspelmeyer}\ \emph {et~al.}(2014)\citenamefont {Aspelmeyer}, \citenamefont {Kippenberg},\ and\ \citenamefont {Marquardt}}]{Aspelmeyer2014}%
  \BibitemOpen
  \bibfield  {author} {\bibinfo {author} {\bibfnamefont {M.}~\bibnamefont {Aspelmeyer}}, \bibinfo {author} {\bibfnamefont {T.~J.}\ \bibnamefont {Kippenberg}}, \ and\ \bibinfo {author} {\bibfnamefont {F.}~\bibnamefont {Marquardt}},\ }\href {\doibase 10.1103/RevModPhys.86.1391} {\bibfield  {journal} {\bibinfo  {journal} {Rev. Mod. Phys.}\ }\textbf {\bibinfo {volume} {86}},\ \bibinfo {pages} {1391} (\bibinfo {year} {2014})}\BibitemShut {NoStop}%
\bibitem [{\citenamefont {Elste}\ \emph {et~al.}(2009)\citenamefont {Elste}, \citenamefont {Girvin},\ and\ \citenamefont {Clerk}}]{Elste2009}%
  \BibitemOpen
  \bibfield  {author} {\bibinfo {author} {\bibfnamefont {F.}~\bibnamefont {Elste}}, \bibinfo {author} {\bibfnamefont {S.~M.}\ \bibnamefont {Girvin}}, \ and\ \bibinfo {author} {\bibfnamefont {A.~A.}\ \bibnamefont {Clerk}},\ }\href {\doibase 10.1103/PhysRevLett.102.207209} {\bibfield  {journal} {\bibinfo  {journal} {Phys. Rev. Lett.}\ }\textbf {\bibinfo {volume} {102}},\ \bibinfo {pages} {207209} (\bibinfo {year} {2009})}\BibitemShut {NoStop}%
\bibitem [{\citenamefont {Li}\ \emph {et~al.}(2019)\citenamefont {Li}, \citenamefont {Korobko}, \citenamefont {Ma}, \citenamefont {Schnabel},\ and\ \citenamefont {Chen}}]{li2019a}%
  \BibitemOpen
  \bibfield  {author} {\bibinfo {author} {\bibfnamefont {X.}~\bibnamefont {Li}}, \bibinfo {author} {\bibfnamefont {M.}~\bibnamefont {Korobko}}, \bibinfo {author} {\bibfnamefont {Y.}~\bibnamefont {Ma}}, \bibinfo {author} {\bibfnamefont {R.}~\bibnamefont {Schnabel}}, \ and\ \bibinfo {author} {\bibfnamefont {Y.}~\bibnamefont {Chen}},\ }\href@noop {} {\bibfield  {journal} {\bibinfo  {journal} {Physical Review A}\ }\textbf {\bibinfo {volume} {100}},\ \bibinfo {pages} {053855} (\bibinfo {year} {2019})}\BibitemShut {NoStop}%
\bibitem [{\citenamefont {Huang}\ and\ \citenamefont {Agarwal}(2017)}]{huang2017}%
  \BibitemOpen
  \bibfield  {author} {\bibinfo {author} {\bibfnamefont {S.}~\bibnamefont {Huang}}\ and\ \bibinfo {author} {\bibfnamefont {G.}~\bibnamefont {Agarwal}},\ }\href@noop {} {\bibfield  {journal} {\bibinfo  {journal} {Physical Review A}\ }\textbf {\bibinfo {volume} {95}},\ \bibinfo {pages} {023844} (\bibinfo {year} {2017})}\BibitemShut {NoStop}%
\bibitem [{\citenamefont {Weiss}\ \emph {et~al.}(2013)\citenamefont {Weiss}, \citenamefont {Bruder},\ and\ \citenamefont {Nunnenkamp}}]{Weiss2013}%
  \BibitemOpen
  \bibfield  {author} {\bibinfo {author} {\bibfnamefont {T.}~\bibnamefont {Weiss}}, \bibinfo {author} {\bibfnamefont {C.}~\bibnamefont {Bruder}}, \ and\ \bibinfo {author} {\bibfnamefont {A.}~\bibnamefont {Nunnenkamp}},\ }\href {http://stacks.iop.org/1367-2630/15/i=4/a=045017} {\bibfield  {journal} {\bibinfo  {journal} {New Journal of Physics}\ }\textbf {\bibinfo {volume} {15}},\ \bibinfo {pages} {045017} (\bibinfo {year} {2013})}\BibitemShut {NoStop}%
\bibitem [{\citenamefont {Weiss}\ and\ \citenamefont {Nunnenkamp}(2013)}]{Weiss2013a}%
  \BibitemOpen
  \bibfield  {author} {\bibinfo {author} {\bibfnamefont {T.}~\bibnamefont {Weiss}}\ and\ \bibinfo {author} {\bibfnamefont {A.}~\bibnamefont {Nunnenkamp}},\ }\href {\doibase 10.1103/PhysRevA.88.023850} {\bibfield  {journal} {\bibinfo  {journal} {Phys. Rev. A}\ }\textbf {\bibinfo {volume} {88}},\ \bibinfo {pages} {023850} (\bibinfo {year} {2013})}\BibitemShut {NoStop}%
\bibitem [{\citenamefont {Kilda}\ and\ \citenamefont {Nunnenkamp}(2016)}]{Kilda2016}%
  \BibitemOpen
  \bibfield  {author} {\bibinfo {author} {\bibfnamefont {D.}~\bibnamefont {Kilda}}\ and\ \bibinfo {author} {\bibfnamefont {A.}~\bibnamefont {Nunnenkamp}},\ }\href {http://stacks.iop.org/2040-8986/18/i=1/a=014007} {\bibfield  {journal} {\bibinfo  {journal} {Journal of Optics}\ }\textbf {\bibinfo {volume} {18}},\ \bibinfo {pages} {014007} (\bibinfo {year} {2016})}\BibitemShut {NoStop}%
\bibitem [{\citenamefont {Vyatchanin}\ and\ \citenamefont {Matsko}(2016)}]{vyatchanin2016}%
  \BibitemOpen
  \bibfield  {author} {\bibinfo {author} {\bibfnamefont {S.~P.}\ \bibnamefont {Vyatchanin}}\ and\ \bibinfo {author} {\bibfnamefont {A.~B.}\ \bibnamefont {Matsko}},\ }\href@noop {} {\bibfield  {journal} {\bibinfo  {journal} {Physical Review A}\ }\textbf {\bibinfo {volume} {93}},\ \bibinfo {pages} {063817} (\bibinfo {year} {2016})}\BibitemShut {NoStop}%
\bibitem [{\citenamefont {Nazmiev}\ and\ \citenamefont {Vyatchanin}(2019)}]{nazmiev2019}%
  \BibitemOpen
  \bibfield  {author} {\bibinfo {author} {\bibfnamefont {A.}~\bibnamefont {Nazmiev}}\ and\ \bibinfo {author} {\bibfnamefont {S.~P.}\ \bibnamefont {Vyatchanin}},\ }\href@noop {} {\bibfield  {journal} {\bibinfo  {journal} {Journal of Physics B: Atomic, Molecular and Optical Physics}\ }\textbf {\bibinfo {volume} {52}},\ \bibinfo {pages} {155401} (\bibinfo {year} {2019})}\BibitemShut {NoStop}%
\bibitem [{\citenamefont {Vostrosablin}\ and\ \citenamefont {Vyatchanin}(2014)}]{Vostrosablin2014}%
  \BibitemOpen
  \bibfield  {author} {\bibinfo {author} {\bibfnamefont {N.}~\bibnamefont {Vostrosablin}}\ and\ \bibinfo {author} {\bibfnamefont {S.~P.}\ \bibnamefont {Vyatchanin}},\ }\href {\doibase 10.1103/PhysRevD.89.062005} {\bibfield  {journal} {\bibinfo  {journal} {Phys. Rev. D}\ }\textbf {\bibinfo {volume} {89}},\ \bibinfo {pages} {062005} (\bibinfo {year} {2014})}\BibitemShut {NoStop}%
\bibitem [{\citenamefont {Tarabrin}\ \emph {et~al.}(2013)\citenamefont {Tarabrin}, \citenamefont {Kaufer}, \citenamefont {Khalili}, \citenamefont {Schnabel},\ and\ \citenamefont {Hammerer}}]{Tarabrin2013}%
  \BibitemOpen
  \bibfield  {author} {\bibinfo {author} {\bibfnamefont {S.~P.}\ \bibnamefont {Tarabrin}}, \bibinfo {author} {\bibfnamefont {H.}~\bibnamefont {Kaufer}}, \bibinfo {author} {\bibfnamefont {F.~Y.}\ \bibnamefont {Khalili}}, \bibinfo {author} {\bibfnamefont {R.}~\bibnamefont {Schnabel}}, \ and\ \bibinfo {author} {\bibfnamefont {K.}~\bibnamefont {Hammerer}},\ }\href {\doibase 10.1103/PhysRevA.88.023809} {\bibfield  {journal} {\bibinfo  {journal} {Phys. Rev. A}\ }\textbf {\bibinfo {volume} {88}},\ \bibinfo {pages} {023809} (\bibinfo {year} {2013})}\BibitemShut {NoStop}%
\bibitem [{\citenamefont {Xuereb}\ \emph {et~al.}(2011)\citenamefont {Xuereb}, \citenamefont {Schnabel},\ and\ \citenamefont {Hammerer}}]{Xuereb2011}%
  \BibitemOpen
  \bibfield  {author} {\bibinfo {author} {\bibfnamefont {A.}~\bibnamefont {Xuereb}}, \bibinfo {author} {\bibfnamefont {R.}~\bibnamefont {Schnabel}}, \ and\ \bibinfo {author} {\bibfnamefont {K.}~\bibnamefont {Hammerer}},\ }\href {\doibase 10.1103/PhysRevLett.107.213604} {\bibfield  {journal} {\bibinfo  {journal} {Phys. Rev. Lett.}\ }\textbf {\bibinfo {volume} {107}},\ \bibinfo {pages} {213604} (\bibinfo {year} {2011})}\BibitemShut {NoStop}%
\bibitem [{\citenamefont {Tagantsev}\ \emph {et~al.}(2018)\citenamefont {Tagantsev}, \citenamefont {Sokolov},\ and\ \citenamefont {Polzik}}]{Tagantsev2018}%
  \BibitemOpen
  \bibfield  {author} {\bibinfo {author} {\bibfnamefont {A.~K.}\ \bibnamefont {Tagantsev}}, \bibinfo {author} {\bibfnamefont {I.~V.}\ \bibnamefont {Sokolov}}, \ and\ \bibinfo {author} {\bibfnamefont {E.~S.}\ \bibnamefont {Polzik}},\ }\href@noop {} {\bibfield  {journal} {\bibinfo  {journal} {Phys. Rev. A}\ }\textbf {\bibinfo {volume} {97}},\ \bibinfo {pages} {063820} (\bibinfo {year} {2018})}\BibitemShut {NoStop}%
\bibitem [{\citenamefont {Tagantsev}\ and\ \citenamefont {Fedorov}(2019)}]{tagantsev2019}%
  \BibitemOpen
  \bibfield  {author} {\bibinfo {author} {\bibfnamefont {A.~K.}\ \bibnamefont {Tagantsev}}\ and\ \bibinfo {author} {\bibfnamefont {S.~A.}\ \bibnamefont {Fedorov}},\ }\href@noop {} {\bibfield  {journal} {\bibinfo  {journal} {Physical review letters}\ }\textbf {\bibinfo {volume} {123}},\ \bibinfo {pages} {043602} (\bibinfo {year} {2019})}\BibitemShut {NoStop}%
\bibitem [{\citenamefont {Mehmood}\ \emph {et~al.}(2019)\citenamefont {Mehmood}, \citenamefont {Qamar},\ and\ \citenamefont {Qamar}}]{mehmood2019}%
  \BibitemOpen
  \bibfield  {author} {\bibinfo {author} {\bibfnamefont {A.}~\bibnamefont {Mehmood}}, \bibinfo {author} {\bibfnamefont {S.}~\bibnamefont {Qamar}}, \ and\ \bibinfo {author} {\bibfnamefont {S.}~\bibnamefont {Qamar}},\ }\href@noop {} {\bibfield  {journal} {\bibinfo  {journal} {Physica Scripta}\ }\textbf {\bibinfo {volume} {94}},\ \bibinfo {pages} {095502} (\bibinfo {year} {2019})}\BibitemShut {NoStop}%
\bibitem [{\citenamefont {Khalili}\ \emph {et~al.}(2016)\citenamefont {Khalili}, \citenamefont {Tarabrin}, \citenamefont {Hammerer},\ and\ \citenamefont {Schnabel}}]{Khalili2016}%
  \BibitemOpen
  \bibfield  {author} {\bibinfo {author} {\bibfnamefont {F.~Y.}\ \bibnamefont {Khalili}}, \bibinfo {author} {\bibfnamefont {S.~P.}\ \bibnamefont {Tarabrin}}, \bibinfo {author} {\bibfnamefont {K.}~\bibnamefont {Hammerer}}, \ and\ \bibinfo {author} {\bibfnamefont {R.}~\bibnamefont {Schnabel}},\ }\href {\doibase 10.1103/PhysRevA.94.013844} {\bibfield  {journal} {\bibinfo  {journal} {Phys. Rev. A}\ }\textbf {\bibinfo {volume} {94}},\ \bibinfo {pages} {013844} (\bibinfo {year} {2016})}\BibitemShut {NoStop}%
\bibitem [{\citenamefont {Huang}\ and\ \citenamefont {Chen}(2018)}]{huang2018}%
  \BibitemOpen
  \bibfield  {author} {\bibinfo {author} {\bibfnamefont {S.}~\bibnamefont {Huang}}\ and\ \bibinfo {author} {\bibfnamefont {A.}~\bibnamefont {Chen}},\ }\href@noop {} {\bibfield  {journal} {\bibinfo  {journal} {Physical Review A}\ }\textbf {\bibinfo {volume} {98}},\ \bibinfo {pages} {063818} (\bibinfo {year} {2018})}\BibitemShut {NoStop}%
\bibitem [{\citenamefont {Huang}\ \emph {et~al.}(2019)\citenamefont {Huang}, \citenamefont {Deng}, \citenamefont {Tan},\ and\ \citenamefont {Cheng}}]{huang2019gen}%
  \BibitemOpen
  \bibfield  {author} {\bibinfo {author} {\bibfnamefont {G.}~\bibnamefont {Huang}}, \bibinfo {author} {\bibfnamefont {W.}~\bibnamefont {Deng}}, \bibinfo {author} {\bibfnamefont {H.}~\bibnamefont {Tan}}, \ and\ \bibinfo {author} {\bibfnamefont {G.}~\bibnamefont {Cheng}},\ }\href@noop {} {\bibfield  {journal} {\bibinfo  {journal} {Physical Review A}\ }\textbf {\bibinfo {volume} {99}},\ \bibinfo {pages} {043819} (\bibinfo {year} {2019})}\BibitemShut {NoStop}%
\bibitem [{\citenamefont {Mehmood}\ \emph {et~al.}(2018)\citenamefont {Mehmood}, \citenamefont {Qamar},\ and\ \citenamefont {Qamar}}]{mehmood2018}%
  \BibitemOpen
  \bibfield  {author} {\bibinfo {author} {\bibfnamefont {A.}~\bibnamefont {Mehmood}}, \bibinfo {author} {\bibfnamefont {S.}~\bibnamefont {Qamar}}, \ and\ \bibinfo {author} {\bibfnamefont {S.}~\bibnamefont {Qamar}},\ }\href@noop {} {\bibfield  {journal} {\bibinfo  {journal} {Physical Review A}\ }\textbf {\bibinfo {volume} {98}},\ \bibinfo {pages} {053841} (\bibinfo {year} {2018})}\BibitemShut {NoStop}%
\bibitem [{\citenamefont {Dumont}\ \emph {et~al.}(2019)\citenamefont {Dumont}, \citenamefont {Bernard}, \citenamefont {Reinhardt}, \citenamefont {Kato}, \citenamefont {Ruf},\ and\ \citenamefont {Sankey}}]{dumont2019}%
  \BibitemOpen
  \bibfield  {author} {\bibinfo {author} {\bibfnamefont {V.}~\bibnamefont {Dumont}}, \bibinfo {author} {\bibfnamefont {S.}~\bibnamefont {Bernard}}, \bibinfo {author} {\bibfnamefont {C.}~\bibnamefont {Reinhardt}}, \bibinfo {author} {\bibfnamefont {A.}~\bibnamefont {Kato}}, \bibinfo {author} {\bibfnamefont {M.}~\bibnamefont {Ruf}}, \ and\ \bibinfo {author} {\bibfnamefont {J.~C.}\ \bibnamefont {Sankey}},\ }\href@noop {} {\bibfield  {journal} {\bibinfo  {journal} {Optics Express}\ }\textbf {\bibinfo {volume} {27}},\ \bibinfo {pages} {25731} (\bibinfo {year} {2019})}\BibitemShut {NoStop}%
\bibitem [{\citenamefont {Tagantsev}(2020{\natexlab{a}})}]{tagantsev2020a}%
  \BibitemOpen
  \bibfield  {author} {\bibinfo {author} {\bibfnamefont {A.~K.}\ \bibnamefont {Tagantsev}},\ }\href@noop {} {\bibfield  {journal} {\bibinfo  {journal} {Physical Review A}\ }\textbf {\bibinfo {volume} {101}},\ \bibinfo {pages} {063813} (\bibinfo {year} {2020}{\natexlab{a}})}\BibitemShut {NoStop}%
\bibitem [{\citenamefont {Tagantsev}(2020{\natexlab{b}})}]{tagantsev2020b}%
  \BibitemOpen
  \bibfield  {author} {\bibinfo {author} {\bibfnamefont {A.~K.}\ \bibnamefont {Tagantsev}},\ }\href@noop {} {\bibfield  {journal} {\bibinfo  {journal} {Physical Review A}\ }\textbf {\bibinfo {volume} {102}},\ \bibinfo {pages} {043520} (\bibinfo {year} {2020}{\natexlab{b}})}\BibitemShut {NoStop}%
\bibitem [{\citenamefont {Tagantsev}\ and\ \citenamefont {Polzik}(2021)}]{tagantsev2021}%
  \BibitemOpen
  \bibfield  {author} {\bibinfo {author} {\bibfnamefont {A.~K.}\ \bibnamefont {Tagantsev}}\ and\ \bibinfo {author} {\bibfnamefont {E.~S.}\ \bibnamefont {Polzik}},\ }\href@noop {} {\bibfield  {journal} {\bibinfo  {journal} {Physical Review A}\ }\textbf {\bibinfo {volume} {103}},\ \bibinfo {pages} {063503} (\bibinfo {year} {2021})}\BibitemShut {NoStop}%
\bibitem [{\citenamefont {Baraillon}\ \emph {et~al.}(2020)\citenamefont {Baraillon}, \citenamefont {Taurel}, \citenamefont {Labeye},\ and\ \citenamefont {Duraffourg}}]{baraillon2020}%
  \BibitemOpen
  \bibfield  {author} {\bibinfo {author} {\bibfnamefont {J.}~\bibnamefont {Baraillon}}, \bibinfo {author} {\bibfnamefont {B.}~\bibnamefont {Taurel}}, \bibinfo {author} {\bibfnamefont {P.}~\bibnamefont {Labeye}}, \ and\ \bibinfo {author} {\bibfnamefont {L.}~\bibnamefont {Duraffourg}},\ }\href@noop {} {\bibfield  {journal} {\bibinfo  {journal} {Physical Review A}\ }\textbf {\bibinfo {volume} {102}},\ \bibinfo {pages} {033509} (\bibinfo {year} {2020})}\BibitemShut {NoStop}%
\bibitem [{\citenamefont {Li}\ \emph {et~al.}(2009)\citenamefont {Li}, \citenamefont {Pernice},\ and\ \citenamefont {Tang}}]{Li2009}%
  \BibitemOpen
  \bibfield  {author} {\bibinfo {author} {\bibfnamefont {M.}~\bibnamefont {Li}}, \bibinfo {author} {\bibfnamefont {W.~H.~P.}\ \bibnamefont {Pernice}}, \ and\ \bibinfo {author} {\bibfnamefont {H.~X.}\ \bibnamefont {Tang}},\ }\href {\doibase 10.1103/PhysRevLett.103.223901} {\bibfield  {journal} {\bibinfo  {journal} {Phys. Rev. Lett.}\ }\textbf {\bibinfo {volume} {103}},\ \bibinfo {pages} {223901} (\bibinfo {year} {2009})}\BibitemShut {NoStop}%
\bibitem [{\citenamefont {Sawadsky}\ \emph {et~al.}(2015)\citenamefont {Sawadsky}, \citenamefont {Kaufer}, \citenamefont {Nia}, \citenamefont {Tarabrin}, \citenamefont {Khalili}, \citenamefont {Hammerer},\ and\ \citenamefont {Schnabel}}]{Sawadsky2015}%
  \BibitemOpen
  \bibfield  {author} {\bibinfo {author} {\bibfnamefont {A.}~\bibnamefont {Sawadsky}}, \bibinfo {author} {\bibfnamefont {H.}~\bibnamefont {Kaufer}}, \bibinfo {author} {\bibfnamefont {R.~M.}\ \bibnamefont {Nia}}, \bibinfo {author} {\bibfnamefont {S.~P.}\ \bibnamefont {Tarabrin}}, \bibinfo {author} {\bibfnamefont {F.~Y.}\ \bibnamefont {Khalili}}, \bibinfo {author} {\bibfnamefont {K.}~\bibnamefont {Hammerer}}, \ and\ \bibinfo {author} {\bibfnamefont {R.}~\bibnamefont {Schnabel}},\ }\href {\doibase 10.1103/PhysRevLett.114.043601} {\bibfield  {journal} {\bibinfo  {journal} {Phys. Rev. Lett.}\ }\textbf {\bibinfo {volume} {114}},\ \bibinfo {pages} {043601} (\bibinfo {year} {2015})}\BibitemShut {NoStop}%
\bibitem [{\citenamefont {Tsvirkun}\ \emph {et~al.}(2015)\citenamefont {Tsvirkun}, \citenamefont {Surrente}, \citenamefont {Raineri}, \citenamefont {Beaudoin}, \citenamefont {Raj}, \citenamefont {Sagnes}, \citenamefont {Robert-Philip},\ and\ \citenamefont {Braive}}]{tsvirkun2015}%
  \BibitemOpen
  \bibfield  {author} {\bibinfo {author} {\bibfnamefont {V.}~\bibnamefont {Tsvirkun}}, \bibinfo {author} {\bibfnamefont {A.}~\bibnamefont {Surrente}}, \bibinfo {author} {\bibfnamefont {F.}~\bibnamefont {Raineri}}, \bibinfo {author} {\bibfnamefont {G.}~\bibnamefont {Beaudoin}}, \bibinfo {author} {\bibfnamefont {R.}~\bibnamefont {Raj}}, \bibinfo {author} {\bibfnamefont {I.}~\bibnamefont {Sagnes}}, \bibinfo {author} {\bibfnamefont {I.}~\bibnamefont {Robert-Philip}}, \ and\ \bibinfo {author} {\bibfnamefont {R.}~\bibnamefont {Braive}},\ }\href@noop {} {\bibfield  {journal} {\bibinfo  {journal} {Scientific reports}\ }\textbf {\bibinfo {volume} {5}},\ \bibinfo {pages} {16526} (\bibinfo {year} {2015})}\BibitemShut {NoStop}%
\bibitem [{\citenamefont {Wu}\ \emph {et~al.}(2014)\citenamefont {Wu}, \citenamefont {Hryciw}, \citenamefont {Healey}, \citenamefont {Lake}, \citenamefont {Jayakumar}, \citenamefont {Freeman}, \citenamefont {Davis},\ and\ \citenamefont {Barclay}}]{Wu2014}%
  \BibitemOpen
  \bibfield  {author} {\bibinfo {author} {\bibfnamefont {M.}~\bibnamefont {Wu}}, \bibinfo {author} {\bibfnamefont {A.~C.}\ \bibnamefont {Hryciw}}, \bibinfo {author} {\bibfnamefont {C.}~\bibnamefont {Healey}}, \bibinfo {author} {\bibfnamefont {D.~P.}\ \bibnamefont {Lake}}, \bibinfo {author} {\bibfnamefont {H.}~\bibnamefont {Jayakumar}}, \bibinfo {author} {\bibfnamefont {M.~R.}\ \bibnamefont {Freeman}}, \bibinfo {author} {\bibfnamefont {J.~P.}\ \bibnamefont {Davis}}, \ and\ \bibinfo {author} {\bibfnamefont {P.~E.}\ \bibnamefont {Barclay}},\ }\href {\doibase 10.1103/PhysRevX.4.021052} {\bibfield  {journal} {\bibinfo  {journal} {Phys. Rev. X}\ }\textbf {\bibinfo {volume} {4}},\ \bibinfo {pages} {021052} (\bibinfo {year} {2014})}\BibitemShut {NoStop}%
\bibitem [{\citenamefont {Meyer}\ \emph {et~al.}(2016)\citenamefont {Meyer}, \citenamefont {Breyer},\ and\ \citenamefont {Köhl}}]{meyer2016}%
  \BibitemOpen
  \bibfield  {author} {\bibinfo {author} {\bibfnamefont {H.~M.}\ \bibnamefont {Meyer}}, \bibinfo {author} {\bibfnamefont {M.}~\bibnamefont {Breyer}}, \ and\ \bibinfo {author} {\bibfnamefont {M.}~\bibnamefont {Köhl}},\ }\href {\doibase 10.1007/s00340-016-6564-z} {\bibfield  {journal} {\bibinfo  {journal} {Applied Physics B}\ }\textbf {\bibinfo {volume} {122}},\ \bibinfo {pages} {290} (\bibinfo {year} {2016})}\BibitemShut {NoStop}%
\bibitem [{\citenamefont {Zhang}\ \emph {et~al.}(2014)\citenamefont {Zhang}, \citenamefont {Barnard}, \citenamefont {McEuen},\ and\ \citenamefont {Lipson}}]{zhang_2014}%
  \BibitemOpen
  \bibfield  {author} {\bibinfo {author} {\bibfnamefont {M.}~\bibnamefont {Zhang}}, \bibinfo {author} {\bibfnamefont {A.}~\bibnamefont {Barnard}}, \bibinfo {author} {\bibfnamefont {P.~L.}\ \bibnamefont {McEuen}}, \ and\ \bibinfo {author} {\bibfnamefont {M.}~\bibnamefont {Lipson}},\ }in\ \href {\doibase 10.1364/CLEO_QELS.2014.FTu2B.1} {\emph {\bibinfo {booktitle} {Proceedings of {CLEO}: 2014, San Jose, CA, 2014}}}\ (\bibinfo  {publisher} {Optical Society of America, San Jose},\ \bibinfo {year} {2014})\ p.\ \bibinfo {pages} {FTu2B.1}\BibitemShut {NoStop}%
\bibitem [{\citenamefont {Buonanno}\ and\ \citenamefont {Chen}(2003)}]{Buonanno2003}%
  \BibitemOpen
  \bibfield  {author} {\bibinfo {author} {\bibfnamefont {A.}~\bibnamefont {Buonanno}}\ and\ \bibinfo {author} {\bibfnamefont {Y.}~\bibnamefont {Chen}},\ }\href {\doibase 10.1103/PhysRevD.67.062002} {\bibfield  {journal} {\bibinfo  {journal} {Phys. Rev. D}\ }\textbf {\bibinfo {volume} {67}},\ \bibinfo {pages} {062002} (\bibinfo {year} {2003})}\BibitemShut {NoStop}%
\bibitem [{\citenamefont {Danilishin}\ and\ \citenamefont {Khalili}(2012)}]{Danilishin2012}%
  \BibitemOpen
  \bibfield  {author} {\bibinfo {author} {\bibfnamefont {S.~L.}\ \bibnamefont {Danilishin}}\ and\ \bibinfo {author} {\bibfnamefont {F.~Y.}\ \bibnamefont {Khalili}},\ }\href {\doibase 10.12942/lrr-2012-5} {\bibfield  {journal} {\bibinfo  {journal} {Living Reviews in Relativity}\ }\textbf {\bibinfo {volume} {15}},\ \bibinfo {pages} {5} (\bibinfo {year} {2012})}\BibitemShut {NoStop}%
\bibitem [{\citenamefont {Walls}\ and\ \citenamefont {Milborn}(2008)}]{Walls2008}%
  \BibitemOpen
  \bibfield  {author} {\bibinfo {author} {\bibfnamefont {D.~F.}\ \bibnamefont {Walls}}\ and\ \bibinfo {author} {\bibfnamefont {G.~J.}\ \bibnamefont {Milborn}},\ }\enquote {\bibinfo {title} {Quantum optics},}\ \ (\bibinfo  {publisher} {Springer},\ \bibinfo {year} {2008})\BibitemShut {NoStop}%
\bibitem [{foo()}]{footnote13}%
  \BibitemOpen
  \href@noop {} {\bibinfo  {journal} {In Eq.(9) from Ref. [2] , "-" in front of g should be repleced with "+"}\ }\BibitemShut {NoStop}%
\bibitem [{\citenamefont {Kampen}(1997)}]{Kampen1997}%
  \BibitemOpen
\bibfield  {journal} {  }\bibfield  {author} {\bibinfo {author} {\bibfnamefont {N.~G.}\ \bibnamefont {Kampen}},\ }\href@noop {} {\bibfield  {journal} {\bibinfo  {journal} {Jounal of Molecular Liquids}\ }\textbf {\bibinfo {volume} {71}},\ \bibinfo {pages} {97} (\bibinfo {year} {1997})}\BibitemShut {NoStop}%
\bibitem [{\citenamefont {Collett}\ and\ \citenamefont {Gardiner}(1984)}]{Collet1984}%
  \BibitemOpen
  \bibfield  {author} {\bibinfo {author} {\bibfnamefont {M.~J.}\ \bibnamefont {Collett}}\ and\ \bibinfo {author} {\bibfnamefont {C.~W.}\ \bibnamefont {Gardiner}},\ }\href {\doibase 10.1103/PhysRevA.30.1386} {\bibfield  {journal} {\bibinfo  {journal} {Phys. Rev. A}\ }\textbf {\bibinfo {volume} {30}},\ \bibinfo {pages} {1386} (\bibinfo {year} {1984})}\BibitemShut {NoStop}%
\end{thebibliography}%
\end{document}